\documentclass[submission,copyright]{eptcs}

\usepackage{graphicx}
\usepackage{amsmath} 
\usepackage{algorithmicx}
\usepackage{algpseudocode}

\usepackage{alltt}
\usepackage{tikz}
\usepackage{adjustbox}

\usepackage{hyperref}
\usepackage{fancyvrb}
\usepackage{alltt}
\usepackage{csvsimple}

\title{Interpolant Tree Automata  and their Application in Horn Clause Verification \thanks{The research leading to these results
   has been supported by the EU
  FP7 project 318337, \emph{ENTRA - Whole-Systems Energy Transparency} and  the EU FP7 project 611004, coordination and support action ICT-Energy.}}

\author{Bishoksan Kafle \institute{Roskilde University, Denmark} \email{kafle@ruc.dk}
\and
John P. Gallagher \institute{Roskilde University, Denmark} 
\institute{IMDEA Software Institute, Spain}
 \email{jpg@ruc.dk}
}

\def\titlerunning{Interpolant Tree Automata and their Application in Horn Clause Verification}
\def\authorrunning{Kafle \& Gallagher}

\begin{document}

\maketitle

\begin{abstract}

This paper investigates the combination of abstract interpretation over the domain of convex polyhedra with interpolant tree automata, in an abstraction-refinement scheme for Horn clause verification. These techniques have been previously applied separately, but are combined in a new way in this paper. The role of an interpolant tree automaton is to provide a generalisation of a spurious counterexample during refinement, capturing a possibly infinite set of spurious counterexample traces.  In our approach these traces are then eliminated using a transformation of the Horn clauses.   We compare this approach with two other methods;  one of them uses interpolant tree automata in an algorithm for trace abstraction and refinement, while the other uses abstract interpretation over the domain of convex polyhedra without the generalisation step. Evaluation of the results of experiments on a number of Horn clause verification problems indicates that the combination of interpolant tree automaton with abstract interpretation gives some increase in the power of the verification tool, while sometimes incurring a performance overhead.



%

\end{abstract}

\newcommand{\rahft}{{\sf \textsc{Rahft}}}
\newcommand{\rahit}{{\sf \textsc{Rahit}}}
\newcommand{\integ}{{\sf int}}
\newcommand{\listint}{{\sf listint}}
\newcommand{\other}{{\sf other}}
\newcommand{\true}{\mathsf {true}}
\newcommand{\false}{\mathsf {false}}
\newcommand{\Bin}{{\sf Bin}}
\newcommand{\Dep}{{\sf Dep}}
\newcommand{\g}{{\sf g}}
\newcommand{\nong}{{\sf ng}}
\newcommand{\OL}{{\cal O}}
\newcommand{\M}{{\sf M}}
\newcommand{\R}{{\cal R}}
\newcommand{\A}{\mathcal{A}}

\newcommand{\body}{\mathcal{B}}
\newcommand{\B}{{\cal B}}
\newcommand{\C}{{\cal C}}
\newcommand{\D}{{\cal D}}
\newcommand{\X}{{\cal X}}
\newcommand{\V}{{\cal V}}
\newcommand{\Q}{{\cal Q}}
\newcommand{\F}{{\sf F}}
\newcommand{\N}{{\cal N}}
\newcommand{\Lang}{{\cal L}}
\newcommand{\powerset}{{\cal P}}
\newcommand{\FTA}{{\cal FT\!A}}
\newcommand{\Term}{{\sf Term}}
\newcommand{\Empty}{{\sf empty}}
\newcommand{\nonEmpty}{{\sf nonempty}}
\newcommand{\compl}{{\sf complement}}
\newcommand{\args}{{\sf args}}
\newcommand{\preds}{{\sf preds}}
\newcommand{\gnd}{{\sf gnd}}
\newcommand{\lfp}{{\sf lfp}}
\newcommand{\psharp}{P^{\sharp}}
\newcommand{\minimize}{{\sf minimize}}
\newcommand{\headterms}{\mathsf{headterms}}
\newcommand{\solvebody}{\mathsf{solvebody}}
\newcommand{\solve}{\mathsf{solve}}
\newcommand{\fail}{\mathsf{fail}}
\newcommand{\member}{\mathsf{memb}}
\newcommand{\ground}{\mathsf{ground}}

\newcommand{\raf}{{\sf raf}}
\newcommand{\qa}{{\sf qa}}
\newcommand{\spl}{{\sf split}}

\newcommand{\transitions}{\mathsf{transitions}}
\newcommand{\nonempty}{\mathsf{nonempty}}
\newcommand{\dom}{\mathsf{dom}}

\newcommand{\Args}{\mathsf{Args}}
\newcommand{\id}{\mathsf{id}}
\newcommand{\type}{\tau}
\newcommand{\restrict}{\mathsf{restrict}}
\newcommand{\any}{\top}
\newcommand{\dyn}{\top}
\newcommand{\dettypes}{{\sf dettypes}}
\newcommand{\Atom}{{\sf Atom}}

\newcommand{\vars}{\mathsf{vars}}
\newcommand{\Vars}{\mathsf{Vars}}
\newcommand{\range}{\mathsf{range}}
\newcommand{\varpos}{\mathsf{varpos}}
\newcommand{\varid}{\mathsf{varid}}
\newcommand{\argpos}{\mathsf{argpos}}
\newcommand{\elim}{\mathsf{elim}}
\newcommand{\pred}{\mathsf{pred}}
\newcommand{\predfuncs}{\mathsf{predfuncs}}
\newcommand{\project}{\mathsf{project}}
\newcommand{\reduce}{\mathsf{reduce}}
\newcommand{\positions}{\mathsf{positions}}
\newcommand{\contained}{\preceq}
\newcommand{\equivalent}{\cong}
\newcommand{\unify}{{\it unify}}
\newcommand{\Iff}{{\rm iff}}
\newcommand{\Where}{{\rm where}}
\newcommand{\qmap}{{\sf qmap}}
\newcommand{\fmap}{{\sf fmap}}
\newcommand{\ftable}{{\sf ftable}}
\newcommand{\Qmap}{{\sf Qmap}}
\newcommand{\states}{{\sf states}}
\newcommand{\head}{\tau}
\newcommand{\atomconstraints}{\mathsf{atomconstraints}}
\newcommand{\thresholds}{\mathsf{thresholds}}
\newcommand{\term}{\mathsf{Term}}
\newcommand{\trees}{\mathsf{trees}}
\newcommand{\renames}{\rho_1}
\newcommand{\renameps}{\rho_2}
\newcommand{\predicates}{\mathsf{Predicates}}
\newcommand{\query}{\mathsf{q}}
\newcommand{\ans}{\mathsf{a}}
\newcommand{\trace}{\mathsf{tr}}
\newcommand{\constr}{\mathsf{constr}}
\newcommand{\Iproj}{\mathsf{proj}}
\newcommand{\SAT}{\mathsf{SAT}}
\newcommand{\interpolant}{\mathsf{interpolant}}
\newcommand{\unknown}{?}
\newcommand{\rhs}{{\sf rhs}}
\newcommand{\lhs}{{\sf lhs}}
\newcommand{\unfold}{{\sf unfold}}
\newcommand{\arity}{{\sf ar}}
\newcommand{\AND}{{\sf AND}}

\def\ll{[\![}
\def\rr{]\!]}

\newcommand{\todo}[1]{\textbf{**}\marginpar{\fbox{
\begin{minipage}{\oddsidemargin}
\textsf{{\small#1}}
\end{minipage}
}}}

\newcommand{\sset}[2]{\left\{~#1  \left|
                               \begin{array}{l}#2\end{array}
                          \right.     \right\}}

\newcommand{\qin}{\hspace*{0.15in}}
\newenvironment{SProg}
     {\begin{small}\begin{tt}\begin{tabular}[t]{l}}%
     {\end{tabular}\end{tt}\end{small}}
\def\anno#1{{\ooalign{\hfil\raise.07ex\hbox{\small{\rm #1}}\hfil%
        \crcr\mathhexbox20D}}}

\newtheorem{property}{Property}
\newtheorem{lemma}{Lemma}
\newtheorem{proof}{Proof}
\newtheorem{definition}{Definition}
\newtheorem{example}{Example}
\newtheorem{theorem}{Theorem}

\section{Introduction}
\label{ch7:intro}

In this paper we  combine two existing techniques, namely abstract interpretation  over the domain of convex polyhedra and  interpolant tree automata in a new way for Horn clause verification. Abstract interpretation is a scalable program analysis technique which computes invariants allowing many program properties to be proven, but suffers from false alarms; safe but not provably safe programs may be indistinguishable from unsafe programs. Refinement is considered in this case.  In previous work \cite{kafleG2015horn} we described  an \emph{abstraction-refinement} scheme for Horn clause verification using abstract interpretation and refinement with finite tree automata. In that approach  refinement eliminates a single spurious counterexample in each iteration of the abstraction-refinement loop, using a clause transformation based on a tree automata difference operation. In contrast to that work,  we apply the method of Wang and Jiao
\cite{WangJ2015} for constructing an \emph{interpolant tree automaton} from an infeasible trace.  This generalises the trace of a spurious counterexample, recognising a possibly infinite number of spurious counterexamples, which can then be eliminated in one iteration of  the abstraction-refinement loop.   We combine this construction in the framework of  \cite{kafleG2015horn}. The experimental results on a set of Horn clause verification problems are reported, and  compared with both \cite{kafleG2015horn} and the results of Wang and Jiao
\cite{WangJ2015} using trace abstraction and refinement.
%
%
%

In Section \ref{ch7:prelim} we introduce the key concepts of constrained Horn clauses and finite tree automata.   Section \ref{ch7:intautomata} contains the definitions of interpolants and the construction of a tree interpolant automaton following the techniques of Wang and Jiao
\cite{WangJ2015}.  In Section \ref{ch7:application} we describe our algorithm combining abstract interpretation with tree interpolant automata, including in Section \ref{ch7:experiments} an experimental evaluation and comparison with other approaches.  Finally in Section \ref{ch7:rel} we discuss related work.

\section{Preliminaries}\label{ch7:prelim}

A constrained Horn clause (CHC) is a first order predicate logic formula of the form 
$\forall(\phi \wedge p_1(X_1) \wedge \ldots \wedge p_k(X_k) \rightarrow  p(X))$ ($k \ge 0$), where $\phi$ is a first order logic formula (constraint)  with respect to some background theory and $p_1,\ldots,p_k, p$ are predicate  symbols. We assume (wlog) that $X_i, X$  are (possibly empty) tuples of distinct variables and $\phi$ is expressed in terms of $X_i, ~X$,  which can be achieved by adding equalities to $\phi$.  $p(X)$ is the head of the clause and $\phi \wedge p_1(X_1) \wedge \ldots \wedge p_k(X_k)$ is the body.    There is a distinguished predicate symbol $\mathit{false}$  which is interpreted as $\false$.   Clauses whose head is $\mathit{false}$ are called \emph{integrity constraints}. Following the notation used in constraint logic programming a clause  is usually written as $H \leftarrow \phi, B_1, \ldots, B_k$  where $H, B_1,..., B_k$ stand for atomic formulas (atoms) $p(X), p_1(X_1), ..., p_k(X_k)$. 
A set of CHCs is sometimes called a (constraint logic) program.

An interpretation of a set of CHCs is represented as a set of \emph{constrained facts} of the form $A \leftarrow \phi$ where $A$ is an atom and $ \phi$ is a  formula  with respect to some background theory. $A \leftarrow \phi$ represents a set of ground facts $A\theta$ such that $\phi \theta$ holds in the background theory ($\theta$ is called a grounding substitution). An interpretation that satisfies each clause in $P$  is called a model of $P$. In some works \cite{DBLP:conf/sas/BjornerMR13,McmillanR2013},  a \emph{model} is also called a \emph{solution} and we use them interchangeably in this paper.

\paragraph{Horn clause verification problem.}
Given a set of CHCs $P$, the CHC verification problem is to check whether there exists a model of $P$. It can easily be shown that $P$ has a model if and only if the fact $\mathit{false}$ is not a consequence of $P$.

\begin{figure}[t]
\centering
\begin{BVerbatim}
c1. fib(A, B):- A>=0, A=<1, B=1.
c2. fib(A, B) :- A>1, A2=A-2, 
           A1=A-1, B=B1+B2,  fib(A1,B1), fib(A2,B2).
c3. false:- A>5, B<A, fib(A,B).  
\end{BVerbatim}
\caption{Example CHCs (Fib) defining a Fibonacci function.}
 \label{ch7:exprogram}
\end{figure}

An example set of CHCs, encoding the Fibonacci function is shown   in Figure  \ref{ch7:exprogram}. Since its derivations are trees, it serves as an interesting example from the point of view of  \emph{interpolant tree automata}.


\begin{definition}[Finite tree automaton \cite{Comon}] An FTA  $\A$ is a tuple $(Q, Q_f, \Sigma, \Delta)$, where 
 $Q$ is a finite set of states,  $Q_f \subseteq Q$ is a set of final states, $\Sigma$ is a set of function symbols, and  $\Delta$ is a set of transitions of the form $f(q_1, \ldots, q_n) \rightarrow q$ with $q, q_1, \ldots , q_n \in Q$ and $f \in \Sigma$. 
 We assume that $Q$ and $\Sigma$ are disjoint.
\end{definition}

We assume that each CHC in a program $P$ is associated with an identifier by a mapping $\id_P: P \rightarrow \Sigma$. An identifier (an element of $\Sigma$) is a function symbol whose arity is the same as the number of atoms in the clause body. For instance a clause $p(X) \leftarrow \phi, p_1(X_1), \ldots, p_k(X_k)$ is assigned a function symbol with arity $k$.  As will be seen later, these identifiers are used to build trees that represent \emph{derivations} using the clauses. A set of derivation trees (traces) of a set of atoms of a program $P$ can be abstracted and represented by an FTA. We provide such a construction in Definition \ref{ch7:trace-fta}.

\begin{definition}[Trace FTA for a set of CHCs] 
\label{ch7:trace-fta} 
Let $P$ be a set of CHCs.  Define the trace FTA for $P$ as $\A_P = (Q,Q_f,\Sigma,\Delta)$ where
\begin{itemize}
\item
$Q = \{ p \mid p ~ is ~a  ~predicate ~symbol ~of ~P \} \cup  \{\mathit{false} \} $;
\item
$Q_f = \{ \mathit{false} \}$;
\item
$\Sigma$ is a set of function symbols;
\item
$\Delta = \{c_j(p_1,\ldots,p_k) \rightarrow p \mid  \mathrm{where~} c_j \in \Sigma, ~p(X) \leftarrow \phi, p_1(X_1), \ldots, p_k(X_k) \in  P, ~c_j = \id_P(p(X) \leftarrow \phi, p_1(X_1), \ldots, p_k(X_k) )\}$.
\end{itemize}
\end{definition}
The elements of $\Lang(\A_P)$ are called trace-terms or trace-trees or simply traces of $P$ rooted at $\mathit{false}$. 

\begin{example}

\label{ch7:exautP}
Let $Fib$ be the set of CHCs in Figure \ref{ch7:exprogram}. Let $\id_{Fib}$ map the clauses to identifiers $c_1,c_2,c_3$ respectively. Then $\A_{Fib} = (Q,Q_f,\Sigma,\Delta)$ where:
\[
\begin{array}{lll}
Q &=& \{\mathtt{fib}, \mathtt{false}\}\\
Q_{f} &=& \{ \mathtt{false}\}\\
 \Sigma &=& \{ c_1, c_2,  c_3\}\\
\Delta &=& \{ c_1 \rightarrow \mathtt{fib}, ~c_2(\mathtt{fib}, \mathtt{fib}) \rightarrow \mathtt{fib}, \\
	&&	c_3( \mathtt{fib}) \rightarrow \mathtt{false} \}
\end{array}
\]
\end{example}

\label{ch7:t-fta}
Similarly, we can also construct an FTA representing a single trace. It should be noted that the whole idea of representing program traces by FTAs is to use automata theoretic operations for dealing with program traces, for example, removal of an undesirable trace from a set of program traces.
Let $P$ be a set of CHCs and let $t \in \Lang(\A_P)$.  There exists an FTA $\A_t$ such that $\Lang(\A_t) = \{t\}$. We illustrate the construction with an example. 

\begin{example}[Trace FTA] 

Consider the FTA in Example \ref{ch7:exautP}. Let $t= c_{3}(c_{2}(c_{1}, c_1)) \in \Lang( \A_P)$.  Then $\A_t = (Q,Q_f,\Sigma,\Delta)$ is defined as:

\[
\begin{array}{lll}
Q &=&\{ \mathtt{e_{1}, e_{2}, e_{3},e_{4}} \}\\
Q_{f} &=&  \{ \mathtt{e_{1}}\} \\
 \Sigma &=& \{ c_1, c_2,  c_3, c_4\}\\
\Delta &=& \{ c_1 \rightarrow \mathtt{e_{3}}, ~c_1 \rightarrow \mathtt{e_{4}}, ~c_2(\mathtt{e_{3}, e_{4}}) \rightarrow \mathtt{e_2}, \\
   && ~c_3(\mathtt{e_2}) \rightarrow \mathtt{e_1} \} \\
\end{array}
\]
\noindent
where $\Sigma$ is the same as in $\A_P$ and the states $\mathtt{e_i}$ $(i=1 \ldots 4)$ represent the nodes in the trace-tree, with root node $\mathtt{e_1}$ as the final state.

\end{example}

A trace-term is a representation of a \emph{derivation trees}, called an $\AND$-tree \cite{Stark89,Gallagher-Lafave-Dagstuhl} giving a proof of an atomic formula from a set of CHCs. 
\begin{definition}[$\AND$-tree for a trace term $T(t)$ (adapted from \cite{kafleG2015horn})]
Let $P$ be a set of CHCs and let $t \in \Lang(\A_P)$.  
An $\AND$-tree corresponding to $t$, denote by $T(t)$, is the following labelled tree, where each node of $T(t)$ is labelled by an atom,  a clause identifier and a constraint.
\begin{enumerate}
\item
For each sub-term $c_j(t_1,\ldots,t_k)$ of $t$ there is a corresponding node in $T(t)$ labelled by an atom $p(X)$, an identifier $c_j$ such that $c_j = \id_P(p(X) \leftarrow \phi, p_1(X_1),\ldots,p_k(X_k))$, and a constraint $\phi$; the node's children (if $k>0$) are the nodes corresponding to $t_1,\ldots,t_k$ and are labelled by $p_1(X_1),\ldots,p_k(X_k)$.
\item
The variables in the labels are chosen such that if a node $n$ is labelled by a clause, the local variables in the clause body do not occur outside the subtree rooted at $n$.

\end{enumerate}
We  assume that each node in $T(t)$ is uniquely identified by a natural number. We omit $t$ from $T(t)$ when it is clear from the context.
\end{definition}

\begin{figure}[h!]
  \centering
    \includegraphics[width=1.0 \textwidth, height=50mm]{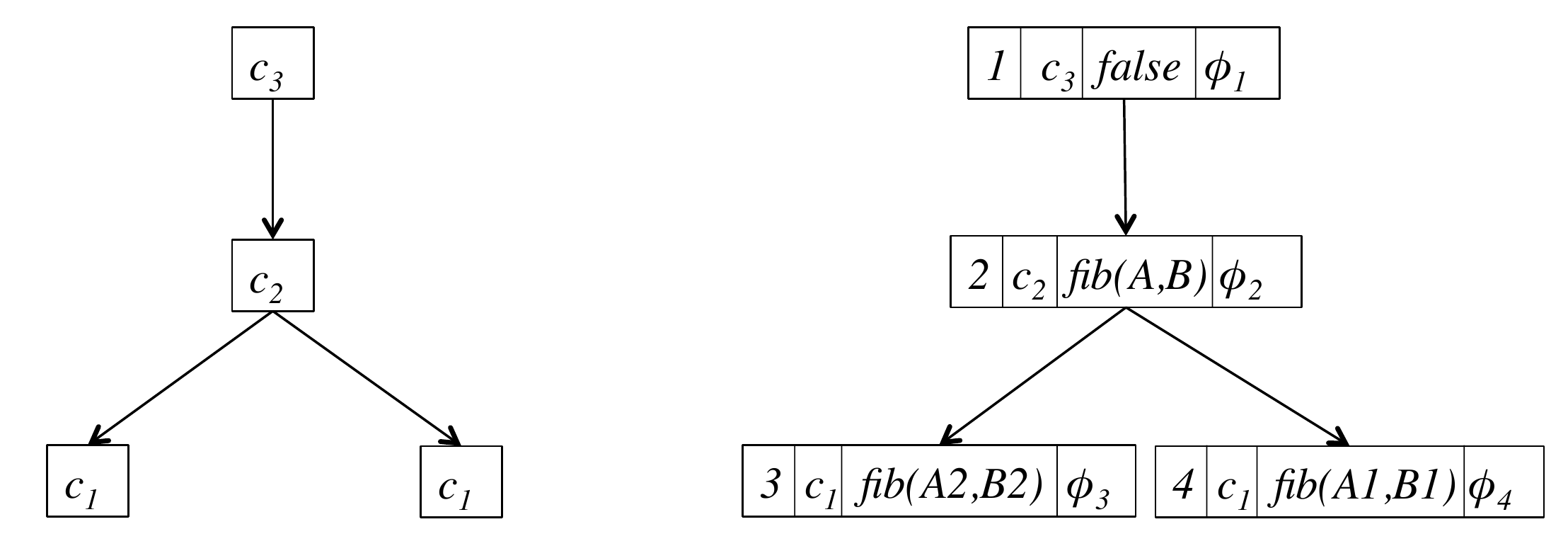}
    \caption{A trace-term $c_3(c_2(c_1,c_1))$ of Fib  (left) and its $\AND$-tree (right), where $\phi_1 \equiv \mathtt{A>5} \wedge \mathtt{B<A}; ~\phi_2 \equiv  \mathtt{A>1} \wedge \mathtt{A2=A-2} \wedge  \mathtt{A1=A-1} \wedge  \mathtt{B=B1+B2};  ~\phi_3 \equiv \mathtt{A2\geq\ 0} \wedge  
\mathtt{A2\leq\ 1} \wedge \mathtt{B2=1}; ~\phi_4 \equiv \mathtt{A1\geq\ 0} \wedge \mathtt{A1\leq\ 1} \wedge \mathtt{B1=1}$.}
   \label{ch7:tracetree}
\end{figure}

The formula represented by an $\AND$-tree $T$, represented by $F(T)$ is 
\begin{enumerate}
\item
$\phi$, if $T$ is a single leaf node labelled by a constraint $\phi$; or
\item
$\phi \wedge  \bigwedge_{i=1..n}(F(T_i))$ if the root node of $T$ is labelled by the constraint $\phi$ and has  subtrees
$T_1,\ldots,T_{n}$.
\end{enumerate}

The formula $F(T)$ where $T$ is the $\AND$-tree in Figure \ref{ch7:tracetree} is 
\[
\begin{array}{l}
\mathtt{A>5} \wedge \mathtt{B<A} \wedge   \mathtt{A>1} \wedge \mathtt{A2=A-2} \wedge  \mathtt{A1= A-1} \wedge \mathtt{B=B1+B2}\\  
 \mathtt{A2\geq\ 0} \wedge \mathtt{A2\leq1} \wedge \mathtt{B2=1} \wedge \mathtt{A1\geq0} \wedge \mathtt{A1\leq1} \wedge \mathtt{B1=1}
\end{array}
\]

We say that an $\AND$-tree $T$ is satisfiable or feasible if $F(T)$ is satisfiable, otherwise unsatisfiable or infeasible. Similarly, we say a  trace-term  is satisfiable (unsatisfiable) iff its corresponding $\AND$-tree is satisfiable (unsatisfiable). The  trace-term $c_3(c_2(c_1,c_1))$ in Figure \ref{ch7:tracetree} is  unsatisfiable since  $F(c_3(c_2(c_1,c_1))) $  is  unsatisfiable.


%
%
%
%

\section{Interpolant tree automata}
\label{ch7:intautomata}

Refinement of trace abstraction is an approach  to program verification \cite{DBLP:conf/sas/HeizmannHP09}. In this approach,  if a property is not provable in an abstraction of program traces then an abstract trace showing the violation of the property is emitted. If such a trace is not feasible with respect to the original program, it is eliminated from the trace abstraction which is viewed as a refinement of the trace abstraction. The notion of interpolant automata  \cite{DBLP:conf/sas/HeizmannHP09} allows one to generalise an infeasible trace to  capture possibly  infinitely many infeasible traces which can then be eliminated in one refinement step. 
In this section, we revisit the construction of an \emph{interpolant tree automaton} \cite{WangJ2015} from an infeasible trace-tree. The automaton serves as a generalisation of the trace-tree; and we apply this construction in Horn clause verification. 

\begin{definition}[(Craig) Interpolant \cite{DBLP:journals/jsyml/Craig57}]  
\label{ch7:interpolant}
Given two formulas $\phi_1, \phi_2 $ such that $\phi_1 \wedge \phi_2 $ is unsatisfiable, a (Craig) interpolant  is a formula $I$ with
(1) $\phi_1 \rightarrow I$; (2) $I \wedge \phi_2 \rightarrow \false$;   and (3) vars($I$) $\subseteq$ vars($\phi_1$) $\cap$ vars($\phi_2$). An interpolant of   $\phi_1$ and $\phi_2$ is represented by $I(\phi_1, \phi_2)$. 
\end{definition}

The existence of an interpolant implies that $\phi_1 \wedge \phi_2$ is unsatisfiable \cite{DBLP:conf/cav/RummerHK13}. Similarly, if the background theory underlying the CHCs $P$ admits (Craig) interpolation \cite{DBLP:journals/jsyml/Craig57}, then every infeasible derivation using the clauses in $P$ has an interpolant \cite{McmillanR2013}.

\begin{example}[Interpolant example]
Let $\phi_1 \equiv \mathtt{A2 \leq 1} \wedge \mathtt{A > 1} \wedge \mathtt{A2 = A - 2} \wedge  \mathtt{A1 = A - 1} \wedge  \mathtt{B = B1 + B2}$ and $\phi_2 \equiv \mathtt{A>5} \wedge \mathtt{B<A}$ such that $\phi_1 \wedge \phi_2$ is unsatisfiable. Since the formula  $I \equiv A \leq 3$  fulfills all the conditions of Definition \ref{ch7:interpolant}, it is an interpolant of $\phi_1$ and $\phi_2$.
\end{example}



Given a node $i$ in an $\AND$-tree $T$, we call $T_i$ the sub-tree rooted at $i$, $\phi_i$  the formula label of node $i$, $F(T_i)$ the  formula of  the sub-tree rooted at node $i$ and $G(T_i)$,  the formula $F(T)$ except  the formula $F(T_i)$, which is  defined as follows: 
\begin{enumerate}

\item 
$\mathit{true}$, if $T$ is a single leaf node labelled by constraint $\phi$ and the node is $i$; or 

\item
$\phi$, if $T$ is a single leaf node labelled by constraint $\phi$  and the node is different from $i$; or

\item
$\mathit{true}$, if the root node of $T$ is labelled by the
constraint $\phi$ and  the node is $i$; or 

\item
$\phi \wedge  \bigwedge_{l=1..n}(G(T_l))$ if the root node of $T$ is labelled by the
constraint $\phi$, and the node is different from $i$ and has  subtrees
$T_1,\ldots,T_{n}$.
\end{enumerate}

\begin{definition}[Tree Interpolant  of an $\AND$-tree (\cite{WangJ2015})] 
\label{ch7:treeInterpolant}
 Let  $T$ be an infeasible $\AND$-tree.    A tree interpolant  $TI(T)$ for $T$ is a tree constructed as follows:
 \begin{enumerate}
 \item 
  The root node $i$ of $TI(T)$ is labelled by  $i$, the atom of the node $i$ of $T$ and the formula  $\mathit{false}$;
 
  \item 
  
 Each leaf node $i$ of $TI(T)$ is labelled by  $i$, the atom of the node $i$ of $T$ and by  $I(F(T_i),  G(T_i))$;
 
 \item 
 Let  $i$ be any other node of $T$. We define $F_1$ as $(\phi_i \wedge \bigwedge_{k=1}^n I_k)$ where $\bigwedge_{k=1}^n I_k$ ($n\geq1$) is the conjunction of formulas representing the interpolants of the children of the node $i$ in $TI(T)$. Then 
the node  $i$ of $TI(T)$ is labelled by  $i$, the atom of the node $i$ of $T$ and the formula $I(F_1 ,  G(T_i))$.
 \end{enumerate}      
\end{definition}

\begin{figure}[h!]
  \centering
    \includegraphics[width=1.0 \textwidth, height=50mm]{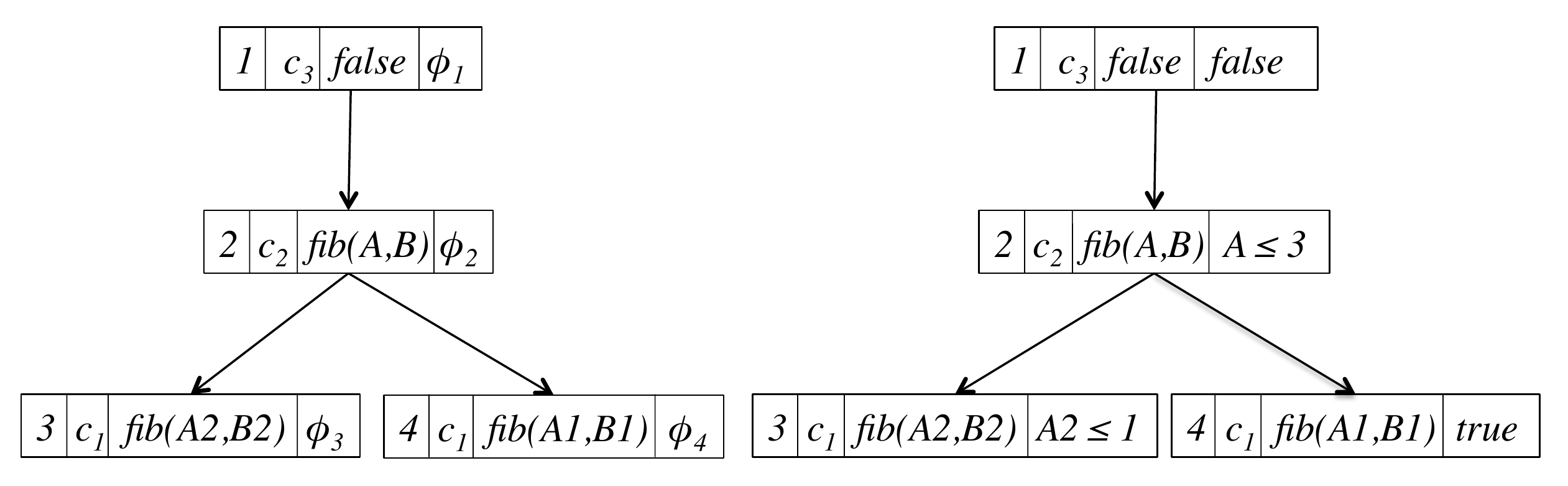}
    \caption{\emph{AND tree}  of Figure \ref{ch7:tracetree}  (left)  and its \emph{tree interpolant} (right). Let $I_j$ represents an interpolant of the node $j$. Then  $I_1 \equiv \mathit{false}$; $I_4 \equiv I(\phi_4, \phi_3 \wedge \phi_1 \wedge \phi_2$) ; $I_3 \equiv I(\phi_3, \phi_1 \wedge \phi_2 \wedge I_4)$;  $I_2 \equiv I(I_3 \wedge I_4 \wedge \phi_2, \phi_1)$.}
     \label{ch7:interpolanttree}
\end{figure}

The \emph{tree interpolant}  corresponding to \emph{AND tree}  in Figure \ref{ch7:tracetree}(b) is shown in Figure \ref{ch7:interpolanttree}(b). 

Since there is a one-one correspondence between an $\AND$-tree and a trace-term, we can define a  tree interpolant for a trace-term as follows:

\begin{definition}[Tree Interpolant of a trace-term $TI(t)$]  Given an infeasible trace-term $t$, its tree interpolant, represented as  $TI(t)$,  is the tree interpolant of its corresponding $\AND$-tree.
\end{definition}

\begin{definition}[Interpolant  mapping $\Pi_{TI}$]  
\label{ch7:interpolantmapping}
Given a tree interpolant  $TI$ for some tree,  $\Pi_{TI}$   is  a mapping from the atom labels  and node numbers of each node in $TI$ to the formula label such that  $\Pi_{TI}(A^j)=\psi$ where $A$ is the atom label  and $\psi$ is the formula label at node $j$. 
\end{definition}
For our example program $\Pi_{TI}$ is the following: 
$$\{ false^1 \mapsto \mathit{false}, fib^2(A,B)  \mapsto A \le 3,   fib^3(A2,B2)  \mapsto A \le 1,  fib^4(A1,B1)  \mapsto \mathit{true} \}$$

\begin{property}[Tree interpolant property]
Let  $TI(T)$ be a tree interpolant  for some infeasible $\AND$-tree $T$. Then 
\begin{enumerate}
\item $\Pi_{TI}(r^i)=\mathit{false}$ where $r$ is the atom label of the root  of $TI(T)$;

\item for each node $j$ with children $j_1, ..., j_n$ ($n\geq 0$) the following property holds: \\
 $(\bigwedge_{k=0}^n \Pi_{TI}(A^{j_k})) \wedge \phi_j \rightarrow \Pi_{TI}(A^{j})$ where $\phi_j$ is the formula label of the node $j$ of $T$;
  
  \item for each node $j$  the following property holds: \\
 $vars(\Pi_{TI}(A^j)) \subseteq (vars(F(T_j)) \cap vars(G(T_j)))$, where the formula $F(T_j)$ and $G(T_j)$ corresponds to $T$.
 
\end{enumerate}
\end{property}

%
%
%
%
%
%
%
%
%
%
%
%
%
%
%
%
%
%
%
%
%
%
%
%
%
%
%
%


\begin{definition}[Interpolant  tree automaton for Horn clauses  $\A^I_t$=$(Q, Q_f, \Sigma, \Delta)$ \cite{WangJ2015}]
\label{ch7:interpolantautomata}
Let $P$ be a set of CHCs, $t \in \Lang(\A_P)$ be any infeasible trace-term and $TI(t)$ be a tree interpolant  of $t$. Let $\sigma: As \times J \rightarrow Q$ where $\sigma$ maps an atom  at node $i \in J$ of $TI(t)$   to an FTA state in $Q$. Define $\rho: Pred^J \rightarrow Pred$ which  maps a predicate name with superscript to a predicate name of $P$. Then the interpolant automaton of $t$  is defined as an FTA  $\A^I_t$ such that

\begin{itemize}
\item
$Q=  \{ \sigma(A,i): A  ~ is  ~ the ~ atom  ~ label  ~ of  ~the  ~ node  ~ i ~  of   ~ TI(t) \}$;
\item
 $F=\{ \sigma(A,i): A  ~ is  ~ the ~ atom ~ label ~of ~the ~root ~of ~ TI(t) \}$; 
\item
$\Sigma$ is a set of function symbols of $P$;
\item
$\Delta = \{c(p_1^{j_1},\ldots, p_k^{j_k}) \rightarrow p^{j} \mid    ~cl = p(X)  \leftarrow \phi, p_1(X_1), \ldots, p_k(X_k) \in P,   ~c = \id_P(cl), ~ \rho(p^i)=p,$ \\
$\rho(p_m^i)=p_m ~for ~m =1..k ~and  ~  \Pi_{TI}(p^{j})(X) \leftarrow \phi,\Pi_{TI}(p_1^{j_1})(X_1), \ldots, \Pi_{TI}(p_k^{j_k})(X_k) \}$.
\end{itemize}

\end{definition}

\begin{example}[Interpolant automata for $c_3(c_2(c_1,c_1))$]

\[
\begin{array}{lll}
Q &=& \{\mathtt{fib^2}, \mathtt{fib^3},\mathtt{fib^4},\mathtt{error}\}\\
Q_{f} &=& \{ \mathtt{error}\}\\
 \Sigma &=& \{ c_1, c_2,  c_3\}\\

\Delta &=& \{ c_1 \rightarrow \mathtt{fib^2},
c_1 \rightarrow \mathtt{fib^3},  
c_1 \rightarrow \mathtt{fib^4}, \\ &&
c_2(\mathtt{fib^2},\mathtt{fib^2}) \rightarrow \mathtt{fib^4},
c_2(\mathtt{fib^2},\mathtt{fib^3}) \rightarrow \mathtt{fib^2}, \\ &&
c_2(\mathtt{fib^2},\mathtt{fib^3}) \rightarrow \mathtt{fib^4},
c_2(\mathtt{fib^2},\mathtt{fib^4}) \rightarrow \mathtt{fib^4}, \\ &&
c_2(\mathtt{fib^3},\mathtt{fib^2}) \rightarrow \mathtt{fib^2},
c_2(\mathtt{fib^3},\mathtt{fib^2}) \rightarrow \mathtt{fib^4}, \\ &&
c_2(\mathtt{fib^3},\mathtt{fib^3}) \rightarrow \mathtt{fib^2},
c_2(\mathtt{fib^3},\mathtt{fib^3}) \rightarrow \mathtt{fib^4}, \\ &&
c_2(\mathtt{fib^3},\mathtt{fib^4}) \rightarrow \mathtt{fib^2},
c_2(\mathtt{fib^3},\mathtt{fib^4}) \rightarrow \mathtt{fib^4}, \\ &&
c_2(\mathtt{fib^4},\mathtt{fib^2}) \rightarrow \mathtt{fib^4},
c_2(\mathtt{fib^4},\mathtt{fib^3}) \rightarrow \mathtt{fib^2}, \\ &&
c_2(\mathtt{fib^4},\mathtt{fib^3}) \rightarrow \mathtt{fib^4},
c_2(\mathtt{fib^4},\mathtt{fib^4}) \rightarrow \mathtt{fib^4}, \\ &&
c_3(\mathtt{fib^2}) \rightarrow \mathtt{error},
c_3(\mathtt{fib^3}) \rightarrow \mathtt{error} \}
\end{array}
\]

\end{example}

The construction described in Definition \ref{ch7:interpolantautomata} recognizes only infeasible traces terms of $P$ as stated in Theorem \ref{ch7:soundness}. 

\begin{theorem}[Soundness]
\label{ch7:soundness}
Let $P$ be a set of CHCs and $t \in \Lang(\A_P)$ be any infeasible trace-term. Then the interpolant  automaton  $\A^I_{t}$  recognises only infeasible trace-terms of $P$.
\end{theorem}

\begin{definition}[Conjunctive interpolant mapping]
\label{ch7:conjunctiveInterpolantmapping}
Given an interpolant mapping  $\Pi_{TI}$ of a tree interpolant  $TI$,  we define  
a conjunctive interpolant mapping for an atom label $A$ of any node in $TI$, represented as $\Pi^c_{TI}(A)$,  to be the following formula    $ \Pi^c_{TI}(A)=\bigwedge_{j } \Pi_{TI}(A^j)$, where $j$ ranges over the nodes of $TI$. It is  the conjunction of interpolants of all the nodes of $TI$ with atom label $A$. The conjunctive interpolant mapping  of $TI$ is represented is $ \Pi^c_{TI} = \{ \Pi^c_{TI}(A) \mid ~A ~is ~the ~atom ~label ~of ~TI\}$.

\end{definition}

%
%
%

It is desirable that the \emph{interpolant tree automaton} of a trace $t \in \Lang(\A_P)$ recognizes as many infeasible traces as possible,  in an ideal situation,  all infeasible traces of $P$. This is possible under the condition described in Theorem \ref{ch7:completeness}.

\begin{theorem}[Model and Interpolant Automata ]
\label{ch7:completeness}
Let   $t \in \Lang(\A_P)$ be any infeasible trace-term.  If $\Pi^c_{TI(t)}$
 is a model of $P$, then the interpolant automaton of $t$ recognises all infeasible trace-terms of $P$.
\end{theorem}
%
%
%
%
%
%
%
%

\section{Application to Horn clause verification}
\label{ch7:application}

\begin{figure}[h!]
\begin{center}
\adjustbox{max width=\linewidth}{
\begin{tikzpicture}
\draw[thick,  color=blue] (0.5,0) rectangle (15,6); 
\begin{scope}
 \node at (6,5.1) {\it FTAM -- Finite tree automata manipulator};
\node at (4.5,5.6) {\it AI --Abstract interpretation};
\node at (12.2,5.1) {\it CG -- Clauses generator};

 
\draw[thick,dashed,green] (.6,1.2) rectangle (7.5,4.5);  
 \node at (5.9,4.1) {\it   Abstraction };

\draw[thick,dashed,red] (7.8,1.2) rectangle (14.7,4.5);  
 \node at (12.6,4.1) {\it   Refinement};


\node at (2.4,3.5) {CHC $P$}; 
 \draw[thick][->] (2.2,3.3) -- (2.2,2.5);

\draw[blue] (1.5,1.5) rectangle (2.7,2.5); 
\node at (2.2,2) {AI }; 


\node at (3.9,2.4) { $\A_P^M$}; 
 \draw[thick][->] (2.7,2.1) -- (4.5,2.1);
 
 \node at (3.9,1.8) { $M$};
\draw[thick][->] (2.7,1.6) -- (4.5,1.6);

    \node at (6.1,3.4) {\bf safe };
   \node at (6.4,2.8) {\bf no };
\draw[thick][->] (6.1,2.5) -- (6.1,3.2);

    \node at (6.1,0.3) {\bf unsafe };
   \node at (6.1,1.1) {\bf yes and feasible };
\draw[thick][->] (6.1,1.5) -- (6.1,0.7);

\node at (10.2,2.5) {$\A_P^M$};
 \draw[thick][->] (7.2,2.3) -- (11.2,2.3);
 
 \node at (10.2,1.9) { $\A^I_t$};
\draw[thick, blue][->] (7.2,2) -- (11.2,2);

\node at (10.2,1.4) { $\A_t$};
\draw[thick][->] (7.2,1.6) -- (11.2,1.6);


 \node at (10.5,0.8) {CHC $P_1$ };
 
  \node at (3,0.8) { $P \leftarrow P_1$ };
  
 \draw[thick][-] (13.9,1.5) -- (13.9,0.5);
  \draw[thick][-] (13.9,0.5) -- (2,0.5);
   \draw[thick][->] (2,0.5) -- (2,1.5);
   
  \draw[thick][->] (13.9,3.2) -- (13.9,2.5);
 \node at (14,3.4) {CHC  $P$};

\draw[blue] (4.5,1.5) rectangle (7.2,2.5); 
\node at (5.9,2.3) {$false \in$ M?}; 
 \draw[-] (4.5,2) -- (7.2, 2);
\node at (6.0,1.8) {$t \in \Lang(\A_P^M)$ };  

\draw[blue] (11.2,1.5) rectangle (12.4,2.5); 
\node at (11.8,2) {FTAM};

\node at (12.9,2.4) {$\A'_P$}; 
 \draw[thick][->] (12.4,2.1) -- (13.4,2.1);

\draw[blue] (13.4,1.5) rectangle (14.6,2.5); 
\node at (14,2) {CG}; 
\end{scope}
\end{tikzpicture}

}
\end{center}
\caption{\it Abstraction-refinement scheme in Horn clause verification \cite{kafleG2015horn}. $M$ is an approximation produced as a result of abstract interpretation. $\A'_P$ recognizes all traces in $\Lang(\A_P^M) \setminus \Lang(\A_t)$. }
\label{ch7:fig:toolchain}
\end{figure}
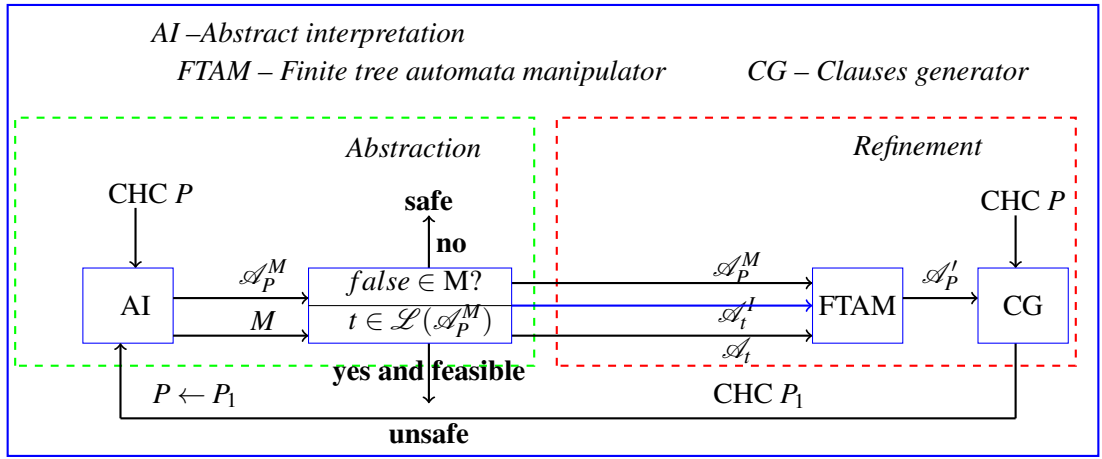

An \emph{abstraction-refinement} scheme for Horn clause verification is described in \cite{kafleG2015horn} which is depicted in Figure \ref{ch7:fig:toolchain}.  In this, a set of CHCs $P$ is analysed using the techniques of \emph{abstract interpretation} over the domain of convex polyhedra which produces an over-approximation $M$ of the minimal model of $P$. The set of traces used during the analysis can be captured by an FTA $\A_P^M$ (see Definition \ref{ch7:model-fta}). This automaton  recognizes all trace-terms of $P$ except some infeasible ones. Some of the infeasible trace-terms are removed by the abstract interpretation.  $P$ is solved or safe (that is, it has a model) if $\mathit{false} \not\in M$. If this is not the case, a trace-term $t \in \Lang(A_P^M)$ is selected and checked for feasibility. If the answer is positive, $P$ has no model, that is, $P$ is unsafe. 

Otherwise $t$ is considered spurious and this drives the refinement process. The refinement in \cite{kafleG2015horn} consists of constructing an automaton $\A'_P$ which recognizes all traces in $\Lang(\A_P^M) \setminus \Lang(\A_t)$ and generating a refined set of clauses from $P$ and  $\A'_P$. The automata difference construction refines a set of traces (abstraction),  which induces refinement in the original program.  The refined program is again fed to the abstract interpreter. This process continues until the problem is safe, unsafe or the resources are exhausted. We call this approach Refinement of Abstraction in Horn clauses using Finite Tree automata, $\rahft$ in short.  

The approach just described lacks generalisation of spurious counterexamples during refinement.  However, in our current approach, we  generalise a spurious counterexample through the use of \emph{interpolant automata}.  Section \ref{ch7:intautomata} describes how to compute an \emph{interpolant  automaton} (taken from \cite{WangJ2015}) corresponding to an infeasible Horn clause derivation. We first construct an \emph{interpolant automaton} viz. $\A^I_t$ corresponding to $t$.  In Figure \ref{ch7:fig:toolchain}, this is shown by a blue line (in the middle) connecting the Abstraction and Refinement boxes. The refinement proceeds as in $\rahft$ with the only difference that $\A'_P$ now recognizes all traces in $\Lang(\A_P^M) \setminus \Lang(\A^I_t)$. We call this approach Refinement of Abstraction in Horn clauses using Interpolant Tree automata, $\rahit$ in short.

Next we briefly describe  how to generate an  FTA, $\A_P^M$, corresponding to a set of clauses $P$ using the approximation produced by \emph{abstract interpretation}. 
Finally we show some experimental results using our current approach on a set of  Horn clause verification benchmarks.

 \paragraph{Obtaining an FTA from a program and a model.}
Let $M$ be a set of constrained atoms of the form $p(X) \leftarrow \phi$ where $p$ is a program predicate and $\phi$ is a constraint over $X$.  
Given such an set $M$, define $\gamma_M$ to be the mapping from atoms to constraints such that $\gamma_M(p(X)) = \phi$ for each constrained fact $p(X) \leftarrow \phi$. $M$ is a model of $P$ (called a \emph{syntactic solution} in \cite{WangJ2015}) if for each clause $p(X) \leftarrow \phi,p_1(X_1), \ldots,p_n(X_n)$ in $P$, $\phi \wedge \bigwedge_{i=1}^n \gamma_M(p_i(X_i)) \rightarrow  \gamma_M(p(X))$.

Given such an $M$, we construct an FTA corresponding to $P$, which is the same as $\A_P$ (Definition \ref{ch7:trace-fta}) except that transitions 
corresponding to clauses whose bodies are not satisfiable in the model are omitted, since they cannot contribute to feasible derivations.

\begin{definition}[FTA defined by a model.]\label{ch7:model-fta}
Let $P$ be a set of CHCs and $M$ be a model defined by a set of constrained facts.  Then the FTA $\A_P^M = (Q,Q_f,\Sigma,\Delta_M)$ where $Q,Q_f$ and $\Sigma$ are the same as for $\A_P$ (Definition \ref{ch7:trace-fta}) and $\Delta_M$ is the following set of transitions.
\[
\begin{array}{ll}
\Delta_M = \{c(p_1,\ldots,p_n) \rightarrow p \mid &id_P(c) = p(X) \leftarrow \phi,p_1(X_1), \ldots,p_n(X_n),\\
& \SAT(\phi \wedge \bigwedge_{i=1}^n \gamma_M(p_i(X_i)))
\end{array}
\]
\end{definition}

\begin{lemma}
Let $P$ be a set of clauses and $M$ be a model of $P$ then $\Lang(\A_P^M)$ includes all feasible trace-terms of $P$ rooted at $\mathit{false}$. 
\end{lemma}
In our experiments, the abstract interpretation was over the domain of convex polyhedra, yielding a set of constrained facts where each constraints is a conjunction of linear equalities and inequalities representing a convex polyhedron.
\begin{example}[FTA produced as a result of \emph{abstract interpretation}]
For our example program in Figure \ref{ch7:exprogram}, the convex polyhedral abstraction produces an over-approximation $M$ which is represented as 
\[
M = \{\mathit{fib(A,B)} \leftarrow  \mathit{A>=0},\mathit{B>=1},\mathit{-A+B>=0}\}
\]
Since there is no constrained fact for $\mathit{false}$ in $M$, this is a model for the example program. Our abstraction-refinement approach terminates at this point. However for the purpose of example, we show the FTA constructed for the example program using $M$. Since the bodies of each clauses except the integrity constraint are satisfied under $M$, the  FTA is same as the one depicted in Example \ref{ch7:exautP} except the transition  $c_3( \mathtt{fib}) \rightarrow \mathtt{false}$, which  is removed because of abstract interpretation.
\end{example}

\subsection{Experiments}
\label{ch7:experiments}

 For our experiment, we have collected a set of 68 programs from different sources.
\begin{enumerate}
\item A set of 30 programs from SV-COMP'15 repository\footnote{http://sv-comp.sosy-lab.org/2015/benchmarks.php} \cite{DBLP:conf/tacas/000115} (recursive category)  and translated them to Horn clauses using inter-procedural encoding of SeaHorn \cite{DBLP:conf/tacas/GurfinkelKN15,DBLP:conf/cav/GurfinkelKKN15} producing  (mostly) non-linear Horn clauses. 

\item A set of 38 problems taken from the source repository\footnote{https://github.com/sosy-lab/sv-benchmarks/tree/master/clauses/LIA/Eldarica}, compiled by the authors of the tool Eldarica \cite{DBLP:conf/fm/HojjatKGIKR12}. This set consists of problems, among others, from  the NECLA static analysis suite, from the paper \cite{DBLP:conf/tacas/JhalaM06}. These tasks  are also considered in \cite{WangJ2015} and  are interpreted over  \emph{integer linear arithmetic}.
\end{enumerate}

We made the following comparison between the tools.

\begin{enumerate}
\item We compare $\rahit$  with  $\rahft$, which compares the effect of removing a set of traces rather than a single trace.  

\item We compare $\rahit$ with the \emph{trace-abstraction} tool \cite{WangJ2015} (\emph{TAR} from now on).   $\rahit$  uses polyhedral approximation combined with trace abstraction refinement  whereas  \emph{TAR} uses only trace abstraction refinement.
\end{enumerate}


The results are summarized in Table~\ref{ch7:tbl:experiments}. 

\paragraph{Implementation:}
Most of the tools in our tool-chain depicted in Figure \ref{ch7:fig:toolchain} are implemented in Ciao Prolog \cite{DBLP:journals/tplp/HermenegildoBCLMMP12} except the one for  determinisation of FTA, which is implemented in Java following the algorithm described in  \cite{DBLP:journals/corr/GallagherAK15}. Our tool-chain obtained by combining  various tools using a \emph{shell script} serves as a proof of concept which is not optimised at all. For handling constraints, we  use the Parma polyhedra library \cite{DBLP:journals/scp/BagnaraHZ08} and the Yices SMT solver \cite{Dutertre:cav2014} over \emph{linear real arithmetic}. The construction of tree interpolation uses constrained based algorithm presented in \cite{DBLP:journals/jsc/RybalchenkoS10} for computing interpolant of two formulas. 

\paragraph{Description:} In Table~\ref{ch7:tbl:experiments},  {\it Program} represents a verification task, {\it Time (secs) RAHFT} and {\it Time (secs) RAHIT} - respectively represent the time in seconds taken by the the tool  $\rahft$ and  $\rahit$ respectively for solving a given task. Similarly,  the number of \emph{abstraction-refinement} iteration needed in these cases to solve a task are represented by   {\it \#Itr. RAHFT} and {\it \#Itr. RAHIT}. Similarly, {\it Time (secs) TAR} and {\it \#Itr. TAR} represent the time taken and the number of iterations  needed by the  tool \emph{TAR}. The experiments were run on a MAC computer running  OS X on  2.3 GHz Intel core i7 processor  and 8 GB memory.

\paragraph{Discussion:}
The comparison between $\rahft$ and $\rahit$ would reflect purely the role of \emph{interpolant tree automata} in Horn clause verification (Table~\ref{ch7:tbl:experiments}) since the only difference between them is the refinement part using \emph{(interpolant) tree automata}. 
The results show that $\rahit$ is more effective in practice than its counterpart $\rahft$. This is justified by the  number of tasks 61/68 solved by $\rahit$ using fewer iterations  compared to $\rahft$, which only solves 56/68 tasks.   This is due to the generalisation of a spurious counterexample during refinement, which also captures other infeasible traces. Since these traces  can be removed in the same iteration,  it (possibly) reduces the number of refinements, however the solving time goes up because of the cost of computing an interpolant automaton.  It is not always the case that $\rahit$ takes less iterations for a task (for example  \emph{Addition03 false-unreach}) than $\rahft$. This is because  the  restructuring of the program obtained  as a result of removing a set of traces may  or may not favour   polyhedral approximation. It is still not clear to us how to produce a right  restructuring which favours  polyhedral approximation.  $\rahit$ times out on \texttt{cggmp2005\_true-unreach}  whereas  $\rahft$ solves it in 5 iterations. We suspect that this is due to the cost of generating  interpolant automata. We are not sure about the complexity of interpolant generation algorithm we used (the size of the formula generated was quite large with respect to the original program, magnitude not known) and there are several calls to the  theorem prover to label each tree node with interpolants.  So the bigger is the trace-tree, the longer it takes to compute the interpolant tree.  In average,  $\rahit$  needs 2.08 iterations and 11.40 seconds time to solve a task whereas $\rahft$ needs 2.32 iterations and 10.55 seconds.

The use of \emph{interpolant tree automata} for trace generalisation and the  tree automata based operations for trace-refinement are same  in both $\rahit$ and  \emph{TAR}. Since \emph{TAR} is not publicly available, we chose the same set of benchmarks  used by \emph{TAR} for the purpose of comparison and presented the results (the results corresponding to \emph{TAR} are taken from \cite{WangJ2015}). The computer used in our experiments and in \emph{TAR} \cite{WangJ2015} have similar characteristics. $\rahit$ solves more than half of the problems  only with \emph{abstract interpretation} over the domain of convex polyhedra without needing any refinement, which indicates its power. $\rahit$ solves 33/38 problems where as \emph{TAR} solves 28/38 problems. In average, $\rahit$ takes less time than \emph{TAR}. In many cases \emph{TAR} solves a task faster than $\rahit$,   however  it spends much longer time in some tasks.  
Our current constraint solver is over   \emph{linear real arithmetic}. If we use it over \emph{linear integer arithmetic} then the results may differ. 
We made some observation with   the problems  \emph{boustrophedon.c},  \emph{boustrophedon\_expansed.c} and \emph{cousot.correct} (which are supposed to be interpreted over integers). In them, if we replace strict inequalities ($>, <$)  with non-strict inequalities ($\geq,\leq$) over integers  (for example replace $X>Y$ with $X \geq Y+1$), then we can solve them only with \emph{abstract interpretation} without refinement which were not solved before the transformation using our solver.  On the other hand, $\rahit$ times out for \emph{ mergesort.error } whereas \emph{TAR} solves it in a single iteration. This indicates that the choice of a spurious counterexample and the quality of interpolant generated from it for generalisation have some effects on verification.

\begin{table}
\resizebox{1.0\textwidth}{106mm}{
    \begin{tabular}{|l|l|l|l|l|l|l|}
    \hline
    \textbf{Program}                              & \textbf{Time (secs) RAHFT} & \textbf{\#Itr. RAHFT} & \textbf{Time (secs) RAHIT} & \textbf{\#Itr. RAHIT} & \textbf{Time (secs) TAR \cite{WangJ2015}} & \textbf{\#Itr. TAR} \\ \hline
    addition                                                 & 1                                     & 0                                          & 1                                     & 0                                          & 0.26                                                            & 3                                        \\ \hline
    anubhav.correct                                          & 2                                     & 0                                          & 2                                     & 0                                          & 1.72                                                            & 9                                        \\ \hline
    bfprt                                                    & 1                                     & 0                                          & 1                                     & 0                                          & 0.43                                                            & 6                                        \\ \hline
    binarysearch                                             & 2                                     & 0                                          & 2                                     & 0                                          & 0.36                                                            & 5                                        \\ \hline
    blast.correct                                            & 5                                     & 1                                          & 11                                    & 1                                          & 8.93                                                            & 65                                       \\ \hline
    boustrophedon.c                                          & TO                                    & -                                          & TO                                    & -                                          & 53.65                                                           & 193                                      \\ \hline
    boustrophedon\_expansed.c                                & TO                                    & -                                          & TO                                    & -                                          & 69.06                                                           & 340                                      \\ \hline
    buildheap                                                & 44                                    & 9                                          & 44                                    & 9                                          & TO                                                              & -                                        \\ \hline
    copy1.error                                              & 11                                    & 0                                          & 11                                    & 0                                          & 12.79                                                           & 19                                       \\ \hline
    countZero                                                & 1                                     & 0                                          & 1                                     & 0                                          & TO                                                              & -                                        \\ \hline
    cousot.correct                                           & TO                                    & -                                          & TO                                    & -                                          & TO                                                              & -                                        \\ \hline
    gopan.c                                                  & 3                                     & 0                                          & 3                                     & 0                                          & TO                                                              & -                                        \\ \hline
    halbwachs.c                                              & TO                                    & -                                          & TO                                    & -                                          & TO                                                              & -                                        \\ \hline
    identity                                                 & 1                                     & 0                                          & 1                                     & 0                                          & 7.67                                                            & 34                                       \\ \hline
    inf1.error                                               & 4                                     & 1                                          & 9                                     & 1                                          & 0.51                                                            & 6                                        \\ \hline
    inf6.correct                                             & 5                                     & 1                                          & 5                                     & 1                                          & 1.96                                                            & 33                                       \\ \hline
    insdel.error                                             & 2                                     & 0                                          & 2                                     & 0                                          & 0.17                                                            & 1                                        \\ \hline
    listcounter.correct                                      & 1                                     & 0                                          & 1                                     & 0                                          & TO                                                              & ~                                        \\ \hline
    listcounter.error                                        & 9                                     & 1                                          & 9                                     & 1                                          & 0.21                                                            & 1                                        \\ \hline
    listreversal.correct                                     & 4                                     & 0                                          & 4                                     & 0                                          & 35.79                                                           & 149                                      \\ \hline
    listreversal.error                                       & 9                                     & 0                                          & 9                                     & 0                                          & 0.3                                                             & 1                                        \\ \hline
    loop.error                                               & 3                                     & 0                                          & 3                                     & 0                                          & 3                                                               & 3                                        \\ \hline
    loop1.error                                              & 8                                     & 0                                          & 8                                     & 0                                          & 10.87                                                           & 19                                       \\ \hline
    mc91.pl                                                  & 139                                   & 24                                         & 7                                     & 3                                          & 0.57                                                            & 7                                        \\ \hline
    merge                                                    & 2                                     & 0                                          & 2                                     & 0                                          & 0.86                                                            & 10                                       \\ \hline
    mergesort.error                                          & TO                                    & -                                          & TO                                    & -                                          & 0.32                                                            & 1                                        \\ \hline
    palindrome                                               & 2                                     & 0                                          & 2                                     & 0                                          & 0.61                                                            & 6                                        \\ \hline
    parity                                                   & 3                                     & 1                                          & 4                                     & 1                                          & 0.62                                                            & 7                                        \\ \hline
    rate\_limiter.c                                          & 3                                     & 0                                          & 3                                     & 0                                          & 49.96                                                           & 130                                      \\ \hline
    remainder                                                & 1                                     & 0                                          & 1                                     & 0                                          & 1.5                                                             & 17                                       \\ \hline
    running                                                  & 3                                     & 1                                          & 8                                     & 2                                          & 0.4                                                             & 5                                        \\ \hline
    scan.error                                               & 3                                     & 0                                          & 3                                     & 0                                          & TO                                                              & -                                        \\ \hline
    string\_concat.error                                     & 6                                     & 0                                          & 6                                     & 0                                          & TO                                                              & -                                        \\ \hline
    string\_concat1.error                                    & TO                                    & -                                          & TO                                    & -                                          & TO                                                              & -                                        \\ \hline
    string\_copy.error                                       & 3                                     & 0                                          & 3                                     & 0                                          & TO                                                              & -                                        \\ \hline
    substring.error                                          & 5                                     & 0                                          & 5                                     & 0                                          & 0.55                                                            & 1                                        \\ \hline
    substring1.error                                         & 15                                    & 0                                          & 15                                    & 0                                          & 2.84                                                            & 5                                        \\ \hline
    triple                                                   & 27                                    & 10                                         & 13                                    & 1                                          & 0.86                                                            & 6                                        \\ \hline
    \hline    \textbf{average (over 38)} & ~                                     & ~                                          & \textbf{8.78}              & \textbf{0.93}                   & \textbf{9.52}                                        & \textbf{38.64}                \\ \hline
    \textbf{solved/total}                         & ~                                     & ~                                          & \textbf{33/38}             & -                                          & \textbf{28/38}                                       & ~                                        \\ \hline
    \hline    Primes\_true-unreach                  & 16                                    & 4                                          & 4                                     & 1                                          & ~                                                               & ~                                        \\ \hline
    sum\_10x0\_false-unreach                                 & 5                                     & 2                                          & 12                                    & 2                                          & ~                                                               & ~                                        \\ \hline
    afterrec\_false-unreach                                  & 2                                     & 1                                          & 3                                     & 1                                          & ~                                                               & ~                                        \\ \hline
    id\_o3\_false-unreach                                    & 6                                     & 3                                          & 7                                     & 3                                          & ~                                                               & ~                                        \\ \hline
    cggmp2005\_variant\_true-unreach                         & 2                                     & 1                                          & 3                                     & 1                                          & ~                                                               & ~                                        \\ \hline
    recHanoi01\_true-unreach                                 & 8                                     & 3                                          & 10                                    & 3                                          & ~                                                               & ~                                        \\ \hline
    cggmp2005b\_true-unreach                                 & 3                                     & 1                                          & 3                                     & 1                                          & ~                                                               & ~                                        \\ \hline
    gcd02\_true-unreach                                      & 11                                    & 4                                          & 11                                    & 4                                          & ~                                                               & ~                                        \\ \hline
    diamond\_false-unreach                                   & 3                                     & 1                                          & 3                                     & 1                                          & ~                                                               & ~                                        \\ \hline
    Addition03\_false-unreach                                & 6                                     & 2                                          & 13                                    & 5                                          & ~                                                               & ~                                        \\ \hline
    diamond\_true-unreach-call1                              & 2                                     & 1                                          & 3                                     & 1                                          & ~                                                               & ~                                        \\ \hline
    id\_i5\_o5\_false-unreach                                & 19                                    & 8                                          & 12                                    & 5                                          & ~                                                               & ~                                        \\ \hline
    diamond\_true-unreach-call2                              & 6                                     & 1                                          & 5                                     & 1                                          & ~                                                               & ~                                        \\ \hline
    cggmp2005\_true-unreach                                  & 10                                    & 5                                          & TO                                    & -                                          & ~                                                               & ~                                        \\ \hline
    gsv2008\_true-unreach                                    & 3                                     & 1                                          & 3                                     & 1                                          & ~                                                               & ~                                        \\ \hline
    Fibocci01\_true-unreach                                  & 52                                    & 10                                         & 29                                    & 6                                          & ~                                                               & ~                                        \\ \hline
    id\_b3\_o2\_false-unreach                                & 5                                     & 2                                          & 3                                     & 1                                          & ~                                                               & ~                                        \\ \hline
    Ackermann02\_false-unreach                               & 68                                    & 17                                         & 25                                    & 7                                          & ~                                                               & ~                                        \\ \hline
    mcmillan2006\_true-unreach                               & 2                                     & 1                                          & 3                                     & 1                                          & ~                                                               & ~                                        \\ \hline
    ddlm2013\_true-unreach                                   & TO                                    & -                                          & 17                                    & 7                                          & ~                                                               & ~                                        \\ \hline
    sum\_2x3\_false-unreach                                  & 2                                     & 1                                          & 3                                     & 1                                          & ~                                                               & ~                                        \\ \hline
    fibo\_5\_true-unreach                                    & TO                                    & -                                          & 77                                    & 7                                          & ~                                                               & ~                                        \\ \hline
    Addition01\_true-unreach                                 & 6                                     & 2                                          & 5                                     & 2                                          & ~                                                               & ~                                        \\ \hline
    Ackermann04\_true-unreach                                & TO                                    & -                                          & 59                                    & 8                                          & ~                                                               & ~                                        \\ \hline
    Addition02\_false-unreach                                & 4                                     & 2                                          & 5                                     & 2                                          & ~                                                               & ~                                        \\ \hline
    id\_i10\_o10\_false-unreach                              & TO                                    & -                                          & 39                                    & 10                                         & ~                                                               & ~                                        \\ \hline
    gcd01\_true-unreach                                      & 9                                     & 4                                          & 5                                     & 2                                          & ~                                                               & ~                                        \\ \hline
    id\_o10\_false-unreach                                   & TO                                    & -                                          & 38                                    & 10                                         & ~                                                               & ~                                        \\ \hline
    gcnr2008\_false-unreach                                  & 13                                    & 4                                          & 6                                     & 2                                          & ~                                                               & ~                                        \\ \hline
    Fibocci04\_false-unreach                                 & TO                                    & -                                          & 91                                    & 11                                         & ~                                                               & ~                                        \\ \hline
    \hline    \textbf{average (over 68)} & \textbf{10.55}             & \textbf{2.32}                   & \textbf{11.40}             & \textbf{2.08}                   & ~                                                               & ~                                        \\ \hline
    \textbf{solved/total}                         & \textbf{56/68}             & ~                                          & \textbf{61/68}             & ~                                          & ~                                                               & ~                                        \\ \hline
    \end{tabular}
   }
   
        \caption{Experiments on  software verification problems. In the table  ``TO'' means  time out which is set for 300 seconds,  ``-'' indicates the insignificance of the result.}
   \label{ch7:tbl:experiments}
\end{table}

\section{Related work}
\label{ch7:rel}

Horn Clauses, as an intermediate language, have become a popular formalism for verification \cite{DBLP:conf/birthday/BjornerGMR15,DBLP:journals/corr/GallagherK14}, attracting both the logic programming and software verification communities \cite{DBLP:journals/corr/BjornerFRS14}. As a result of these,  several verification techniques and tools have been developed for CHCs, among others, \cite{DBLP:conf/cav/GurfinkelKKN15,DBLP:conf/pldi/GrebenshchikovLPR12,DBLP:conf/pepm/KafleG15,DBLP:conf/tacas/AngelisFPP14,kafleG2015horn,DBLP:conf/cav/JaffarMNS12,DBLP:conf/fm/HojjatKGIKR12}. To the best of our knowledge, the use of automata based approach for \emph{abstraction-refinement} of Horn clauses is relatively new \cite{kafleG2015horn,WangJ2015}, though the original framework  proposed for imperative programs goes back to \cite{DBLP:conf/sas/HeizmannHP09,DBLP:conf/popl/HeizmannHP10}.

 The work described in \cite{kafleG2015horn} uses FTA based approach for refining \emph{abstract interpretation} over the domain of convex polyhedra \cite{Cousot-Halbwachs-78}, which  is similar  to \emph{trace abstraction}  \cite{DBLP:conf/sas/HeizmannHP09,DBLP:conf/cav/HeizmannHP13, WangJ2015} with the following differences. In \cite{kafleG2015horn}, there is an interaction between state  abstraction by \emph{abstract interpretation} \cite{DBLP:conf/popl/CousotC77}   and trace abstraction by FTA but there is no generalisation of spurious counterexamples. On one hand,  \cite{DBLP:conf/sas/HeizmannHP09,DBLP:conf/cav/HeizmannHP13, WangJ2015} use  \emph{trace-abstraction} with the generalisation of spurious counterexamples using \emph{interpolant automata} and may diverge from the solution due to the lack of right generalisation.   On the other hand, \emph{abstract interpretation} \cite{DBLP:conf/popl/CousotC77} is one of the most promising  techniques for verification which is scalable but suffers from \emph{false alarms}. When combined with refinement \emph{false alarms} can be minimized.  Our current work takes  the best of both of these approaches.

\section{Conclusion}

 This paper brings together  \emph{abstract interpretation} over the domain of convex polyhedra and \emph{interpolant tree automata} in an \emph{abstraction-refinement} scheme for Horn clause verification  and combines them in a new way. Experimental results on   a set of  software verification benchmarks  using this scheme demonstrated their usefulness in practice; showing  some slight improvements over the previous approaches. In the future, we plan to evaluate its effectiveness in a larger set of benchmarks, compare our approach with other similar approaches  and  improve the implementation aspects of our tool. Further study is needed to find a suitable combination of  abstract interpretation and interpolation based techniques, based on a deeper understanding of the interaction among interpolation, trace elimination and abstract interpretation.

\bibliographystyle{eptcs}
\bibliography{refs}

\end{document}